\magnification=\magstep0
\hsize=13.5 cm
\vsize=19.0 cm
\baselineskip=12 pt plus 1 pt minus 1 pt
\parindent=0.5 cm %
\hoffset=1.3 cm %
\voffset=2.5 cm %

\font\twelvebf=cmbx10 at 12truept
\font\twelverm=cmr10 at 12truept
\overfullrule=0pt
\nopagenumbers

\newtoks\leftheadline \leftheadline={\hfill {\eightit Srabani Datta}
\hfill}
\newtoks\rightheadline \rightheadline= {\hfill {eightit The Structure of
Molecular Clouds}
\hfill}
 \newtoks\firstheadline \firstheadline={{\eightrm Automated Data
Analysis in Astronomy , {\eightbf xx-xx}}\hfill}

\def\makeheadline{\vbox to 0pt{\vskip -22.5pt
\line{\vbox to  8.5 pt {}\ifnum\pageno=1\the\firstheadline
\else%
\ifodd\pageno\the\rightheadline\else%
\the\leftheadline\fi\fi}
\vss}\nointerlineskip}

\font\eightrm=cmr8 \font\eighti=cmmi8 \font\eightsy=cmsy8
\font\eightbf=cmbx8 \font\eighttt=cmtt8 \font\eightit=cmti8
\font\eightsl=cmsl8
\font\sixrm=cmr6  \font\sixi=cmmi6  \font\sixsy=cmsy6
\font\sixbf=cmbx6

\def\eightpoint{\def\rm{\fam0\eightrm}
\textfont0=\eightrm \scriptfont0=\sixrm \scriptscriptfont0=\fiverm
\textfont1=\eighti \scriptfont1=\sixi \scriptscriptfont1=\fivei
\textfont2=\eightsy \scriptfont2=\sixsy \scriptscriptfont2=\fivesy
\textfont3=\tenex \scriptfont3=\tenex \scriptscriptfont3=\tenex
\textfont\itfam=\eightit \def\it{\fam\itfam\eightit}%
\textfont\slfam=\eightsl \def\sl{fam\slfam\eightsl}%
\textfont\ttfam=\eighttt \def\tt{\fam\ttfam\eighttt}%
\textfont\ttfam=\eightbf \scriptfont\bffam=\sixbf
\scriptscriptfont\bffam=\fivebf \def\bf{\fam\bffam\eightbf}%
\normalbaselineskip=10pt plus 0.1 pt minus 0.1 pt
\normalbaselines
\abovedisplayskip=10pt plus 2.4pt minus 7pt
\belowdisplayskip=10pt plus 2.4pt minus 7pt
\belowdisplayshortskip=5.6pt plus 2.4pt minus 3.2pt \rm}

\def\frac#1#2{{#1\over#2}}

\def\pmb#1{\setbox0=\hbox{$#1$}\kern-0.015em\copy0\kern-\wd0%
\kern0.03em\copy0\kern-0.015em\raise0.03em\box0}

\pageno=1
\vglue 60 pt

\leftline{\twelvebf Fractal Structure of Molecular Clouds}

\smallskip
\vskip 46 pt

\leftline {\twelverm Srabani Datta}

\vskip 4pt
\leftline{ Department of Applied Mathematics, Science College,}
\leftline{ University of Calcutta, 92 A.P.C. Road, Calcutta 700 009,
 India} 
\vskip 20 pt

\leftheadline={\hfill{\eightit Srabani Datta}\hfill}
\rightheadline={\hfill{\eightit Structure of Molecular Clouds}\hfill}

{\parindent=0cm\leftskip=1.5cm {\bf Abstract.}

\noindent Compelling evidence exists to show that the structure of 
molecular clouds  is fractal  in nature.  In  this poster,  the author
reiterates that view and, in  addition, asserts that not only is cloud
geometry fractal,  but that they  also have a common  characteristic -
they are similar in shape to the Horsehead in Orion. This shape can be
described by  the Julia function f(z) =  z$^2$ + c,where both  z and c
are complex quantities  and c = -0.745429 +  0.113008i.  The dynamical
processes responsible  for the production  of these clouds seem  to be
turbulence  followed  by  Brownian  motion  till  high  densities  are
reached,   at  which   point  structure   formation  is   dictated  by
gravity. The  author presents image analysis of  four  varied examples,
namely  those  of the  Horsehead  nebula,  Helix nebula,Eagle  nebula,
Rosette nebula and  Paley I nebula to prove her  hypothesis. The images
of fractal  these nebulae are  analyzed for their box  dimension using
fractal  analysis software  and comparisons  are made  with  the given
Julia set.

\smallskip

\vskip 0.5 cm


\vskip 20 pt

\centerline{\bf 1. Introduction }
\bigskip
\noindent
Molecular   clouds   are   self-gravitating,  magnetized,   turbulent,
compressible   fluids  (J.P.Williams   et.al.    1999; Elmegreen  1999;
R.B.Larson  1995).   In addition,  their  molecular structure  is
similar over  a wide range of  scales with similar  power law indices,
quite independant of  the star-forming nature of the  cloud, even at a
resolution of a few parsecs.  The only difference between these clouds
has been found to be in  their average densities in that in regions of
high densities, gravity begins  to dominate and self-similarity breaks
down,  ultimately leading  to the  formation of  stars,  planets, etc.
This  implies  a hierarchy  of  structure,  beginning  from the  giant
molecular clouds with sizes about 50 pc and masses ranging from 10$^4$
to 10$^6$  $M_\odot$.The next  class is the  normal clouds  with sizes
ranging from  a pc  to 30 pc  and masses  from 25 $M_\odot$  to 10$^4$
$M_\odot$ (S.S.Prasad 1995).  These clouds may or may not be bound and
may  contain  a  small  number  of low  mass  stars.   However,  their
contribution  to  the total  star  formation  rate  in the  Galaxy  is
negligible. The third is the clump within the cloud. Clumps have sizes
ranging from 0. 4  to a few pc amd masses from  a few to 25 $M_\odot$.
They are  coherent star-forming regions out of  which stellar clusters
form. The  gas clumps  are normally bound  by pressure rather  than by
gravity.   However, both clumps  bound by  strong self-gravity  and by
external pressure have identical density distributions. The fourth and
final class is that of cores found within clumps, which have sizes 0.1
pc tp  0.4 pc and masses  less than 1 $M_\odot$.   Cores are regions
out  of which  single stars  (or multiple  systems) are  born  and are
always  gravitationally  bound.  Stars  formed  in  cores may  accrete
matter from the host clump or cloud at the protostar stage.

\smallskip

Power-law relationships in molecular clouds have been found in earlier
quantitative   studies   (Kramer  et.    al.1998;   Hetem  \&   Lepine
1993;Stutzki et.al.1998). In this  poster, the author investigates the
structure  of  certain molecular  cloud  images,  by  using a  fractal
dimension analysis software. The images presented are of (i) Horsehead
nebula (B33) in Orion (Kramer  et.  al.  1996), (ii)Eagle nebula (M16)
in Serpens (Hester et.al. 1996), (iii)Rosette nebula
(NGC  2237 to NGC  2246) (Carlqvist  et.al. 1998;Clayton  et.al. 1998;
White 1997) and (iv) IRAS 02356-2959 (Paley I) in Fornax (Paley et.al.
1991; Stark 1995).Of these four, that of the Horsehead nebula has been
obtained  from the  author's observations  from the  2.3m  Vainu Bappu
Telescope,  Kavalur, India.   Basic mathematical  background  for this
work is presented in section 2. , observations in section 3., analysis
in section 4.  Finally, discussions are presented in section 5.

\vskip 20 pt
\centerline{ {\bf 2. Mathematical Background }}

\bigskip

\noindent
Fractals are self-similar objects,  having fine structure. Examples of
fractal objects include the von Koch curve, the Sierpinski gasket, the
Cantor  set  and the  Julia  set(  K.   Falconer 1997;  H.O.   Peitgen
et.al. 1992). In order to study fractal structure in physical objects,
we need a quantitative measure defined on them. Such a quantity is the
principal definition of fractal dimension , other definitions, such as
Kolmogorov (or  Minkowski) dimension and box (or  grid) dimension, are
widely used. Indeed, in practise, box dimension is the easiest to work
with  and  is   equivlent  to  Kolmogorov  dimension.The  box-counting
dimension is  defined in  the following  way : if  X is  any non-empty
bounded subset of R$^n$ and $ N_\epsilon(X)$ be the smallest number of
sets  of diameter  at  most $\epsilon$  which  can cover  X, then  box
dimension is defined as

$$ dim X = \lim_{\epsilon\to 0} log N_\epsilon(X)/log(1/\epsilon) $$

\noindent
In the  above definition of  dimension, the fundamental  assumption is
that  for each  $\epsilon$, we  measure a  set in  such a  manner that
irregularities  of size  less than  $\epsilon$ will  be  ignored.  For
example, if  X is  a plane  curve, then $  N_\epsilon(X)$ will  be the
number  of  steps  required  by  a  pair of  dividers  set  at  length
$\epsilon$ to traverse  X. We have to also keep in  mind that for real
phenomenon we can use only a finite range of $\epsilon$. An example of
a fractal object is that of the Julia set given by the function
$$ f(z) = z^2 + c $$

\noindent 
where z and  c are complex numbers and  c = -0.745429+0.113008i(Fig.1)
However, unlike Euclidean objects,  a fractal object retains a similar
level  of  complexity  for  each  value of  $  \epsilon$,  the  fractal
dimension gives  an index of the  degree to which  a fractal structure
fills the  space in  which it is  embedded (B.Mandelbrot  1982) Hence,
examination of  an object for  self-similarity at different  scales of
magnification can suggest whether the object has a fractal structure.

\vskip 20 pt
\centerline{{\bf 3.Observations}}
\bigskip

\noindent
Images  of  the  Horsehead  nebula   were  obtained  with  the  2.3  m
Cassegranian telescope at the Vainu Bappu Observatory (VBO) located in
Kavalur,  India,  1$^{st}$  and 3$^{rd}$  April 2000.   The SITe
1024x1024 pixels, 24 micron  charge-coupled device (CCD) was placed at
it's prime focus.  The camera projects  each pixel to 0.66" on the sky
(resolution of 27.6" per mm) giving a field of view of 11.264 x 11.264
". The images were obtained  through H-alpha (6563 \AA )(Fig.2), V and
R filter with FWHM of H-alpha filter  being 100 \AA, while the V and R
filters were Bessel filters  with the standard bandpass. The effective
exposures of  the images are given  in Table 1. The  seeing during the
observation of the order of a  few arc seconds.   Each image  was bias
subtracted  and  flat-fielded   using  twilight  flats  obtained  each
day. However, dark currents were  assumed to be neglible as the camera
was cooled to about - 90$^0$  C. Both bad regions and cosmic rays were
removed  using IRAF  NOAO image  reduction software.  The CCD,  with a
read-out noise of the order  of 8 electrons, was calibrated using IRAF
software.   The standard  stars used  for guiding  the  telescope were
obtained from the Hubble Space Telescope (HST) Guide Star Catalogue.

\bigskip

\medskip

\centerline{Table 1. Observation details from the 2.3m VBT, Kavalur.}
\centerline {Co-ordinates are based on year 2000.}
\medskip

\vbox{\tabskip= 0pt \offinterlineskip
\def\tablerule{\noalign{\hrule}}
\halign to \hsize{\strut#&\vrule# \tabskip= 1em plus 2em&
\hfil#& \vrule#& \hfil#& \vrule#& \hfil#& \vrule#& \hfil#& \vrule#&
\hfil#& \vrule#& \hfil#& \vrule# \tabskip= 0pt\cr\tablerule
&&Date&&Right  Acsension&&Declination  &&Filter   &&  UT  &&  Exposure
time&\cr\tablerule

&&2.4.00&&  5h40m58.6s  &&$-2^0 27'  23.84"$&&  H-alpha&&  1357 &&  30
&\cr\tablerule
 &&3.4.00&&  5h40m58.6s &&$-2^0 27' 23.84"$&&  R && 1353
&& 15 &\cr\tablerule
 &&3.4.00&&  5h40m58.6s &&$ -2^0 27' 23.84"$ &&V&&
1426 &&30 &\cr\tablerule
}}

\bigskip

\vskip 3.5cm
\bigskip
\centerline{\bf 4. Fractal Analysis}
\bigskip

\noindent
The images (Figures  3 to 5 ) were obtained  from various sources (HST
website;   D.Malin  1983;R.Stark  1995)   and  converted   to  Windows
bitmaps. This causes the digitized image to be stored as a data matrix
where pixels belonging to the pattern  are stored as 1 and pixels from
the background  are stored  as 0, or  vice versa.  This  procedure can
also  be  carried out  for  colour images  (which  are  overlays of  3
grey-scale  images. The fractal  dimension of  these images  were then
measured  using  an  implementation  of  the  box-counting  method.The
automated  fractal  image  analysis  software,  Benoit  1.3(  Truesoft
International Inc.,St.Petersburg, USA), was  used to convert the solid
images  to  black and  white  pixel  outlines.This  software has  been
reviewed(  Science  1999,285,pg.1228)and  has  been found  to  perform
satisfactorily. These images  were overlayed with grids in  such a way
that the minimum number of  boxes were occupied.The is accomplished in
Benoit  by rotating  the  grid for  each  size through  90 degrees  at
various angular increments of rotation.The  user is able to change the
box length decrease factor as well as the size of the largest box.

\bigskip
\noindent
\midinsert
\vskip 6.0cm
\vskip 0.5cm
{\eightpoint %
\noindent
{\ Figure1.The Julia set for c = -0.745429+ 0.113008i(Box dimension=1.679594)}  }
\endinsert

\vskip .5cm


\bigskip
\noindent
\midinsert
\vskip 3.5 cm

\vskip 0.5cm
{\eightpoint %
\noindent
{\ Figure2.   H-alpha image of  the Horsehead  nebula obtained  with the
2.3m VBT, Kavalur(Box dimension=1.6965725) }
}
\vskip 1.3cm
\endinsert

\bigskip
\noindent 
\midinsert
\vskip 2.5cm
\vskip 0.5cm
{\eightpoint %
{ Figure 3. Part of  PaleyI (Stark 1995)(Box dimension= 1.67718) }
}
\vskip 0.3 cm
\endinsert


\bigskip

\noindent
\midinsert
\vskip 3.5cm
\vskip 0.5cm
{\eightpoint %
{Figure 4. Closeup of a trunk of the Rosette nebula (Malin 1983)(Box dimension=1.78261).}}
\vskip 0.3cm
\endinsert


\bigskip
\noindent
\midinsert
\vskip 3.5cm
\vskip 0.5cm
{\eightpoint %
{ Figure 5. Closeup of a trunk of M16(part of column I in Hester 1995)obtained from the Hubble Picture Gallery(dimension=1.669622)}
}
\vskip 0.3cm
\endinsert

For the analysis,  a set of ten measurements were  taken at fixed grid
rotation increments of 15 degrees. An average dimension was calculated
and the number of iterations  giving the slope with the least standard
deviation  (SD)  while  being  nearest  to  the  to  the  average  was
chosen. Log-log graphs were then plotted of the reciprocal of the side
length  of  the  square  against  the  number  of  outline  containing
squares. The slope is estimated by  fitting a line using the method of
least  squares.Ten further sets  of measurements  were taken  with this
selected number  of iterations and  box size decrease  factor. Results
for  each  cloud were  tabulated  and then  the  Student's  t test  of
significance (Table 2)  was used to test for  their deviation from the
value of  the dimension for  a Euclidean shape(  topological dimension
1). A  further test was made  for fluctuations of  the cloud dimension
values from  that of the  Julia set. The  results show that  the cloud
dimensions are significantly different from it's topological dimension
at  1  percent  cofidence  limit  (t  value 3.25)  but  not  from  the
Kolmogorov dimension  of the  Julia set.  Hence,  it can be  said with
certainity that  molecular clouds are  not only fractal but  also that
their dimension is that of the given Julia set .

\medskip

\centerline{Table 2. Analysis of box dimension .}
\medskip

\vbox{\tabskip=0pt \offinterlineskip
\def\tablerule{\noalign{\hrule}}
\halign to \hsize{\strut#& \vrule#\tabskip=1em plus2em&
\hfil#& \vrule#& \hfil#& \vrule#& \hfil#& \vrule#& \hfil#& \vrule#&
\hfil#& \vrule#\tabskip=0pt\cr\tablerule
&&Name  && Pattern  size(pixels) && Dimension&&Student's t value&&
&\cr\tablerule
  &&Julia   set&&  794x1324  &&   1.679594  &&  91.19396
&\cr\tablerule
 &&Horsehead nebula && 406x473 && 1.6965725&& 16.52918&
\cr\tablerule
&&Paley I && 463x442 && 1.677181 && 95.62187 &
\cr\tablerule
&&   Rosette  nebula(close-up)&&   109x81  &&   1.78261   &&  85.69697
&\cr\tablerule 
 && Eagle  nebula(close-up) &&  188x212 &&  1.669622 &&
70.454 &\cr\tablerule
 }}
\bigskip

\vskip 20 pt
\centerline{\bf 5. Discussion}
\bigskip

The  discovery  of  power-law  relationships in  molecular  clouds  by
Larson(1981)   was   a   ground-   breaking   study   in   favour   of
self-similarity, especially  so since a power-law can  only be applied
to  clouds with masses  greater than  10$^4$ $  M_\odot$ and  does not
apply  to cores  and gravitationally  bound  clumps. My  study on  the
different types of nebulae  reinforces this view, especially since the
parts studied are small sections of the main trunk in some clouds (eg.
Eagle and Rosette).  What is  significant here is that their dimension
is the dimension  of the Julia set. This  result implies that physical
features  of  clouds  can  be  explained  if  we  make  the  following
assumptions(i) cloud  structure resembles the  Julia set corresponding
to the Mandelbrot  constant given by c =  -0.745429+0.113008i and that
it is a dynamical entity  composed of cloud matter, (ii) growth occurs
around the  central point  (an attractor) of  each spiral making  it a
potential site of star formation,  (iii) groups of such central points
spiral  inwards  towards a  massive  central  point, identifying  such
central points  to be dark  objects.  Such central objects  may become
supermassive  black holes  (Macchetto  et.  al.   1997)  or remain  as
numerous small  objects( Goodman and Lee 1989),  perhaps even composed
of strange  matter (van  der Marel 1997;  Weber 2000; Chandra \& Goyal
2000) (iv) central  points increase in size  as we move  away from the
massive central attractor, (v) growth also occurs along each spiral of
the set in  the form of horseheads, whose  sizes reduce adinfinitum as
the spiral moves  towards the attractive fixed point  and finally (iv)
after  the  elapse  of  sufficient  time,  when  adequate  matter  has
accumulated in  the centre, a  dwarf, a star  or group of  stars would
form, depending on the size of the clump.

\bigskip
\centerline{\bf References}
\bigskip
{\eightpoint\parindent=0pt\everypar={\hangindent=0.5 cm}

Carlqvist, P., et. al., 1998, Astron. Astrophys. 332, L5.

Chandra, D. \& Goyal, A.,2000, Phys. Rev. D, in press.

Clayton, A., et. al., 1999, Astron. Astrophys. 334, 264.

Elmegreen, B.G., 1999, Ap.J 527, 266.

Falconer,K.,1997,Fractal Geometry,Wiley and Sons, Chichester .

Goodman,J. and Lee, H.M. ,1989, Ap.J. 337, 84.

Hester, J.J., et.al., 1996, A.J. 111, 2349.

Hetem, A.jr. and Lepine, J.R.D., 1993,Astron. Astrophys.270, 451.

Kramer,       C.;       Stutzki,J.and      Winnewisser,G.;       1996;
Astron.Astrophs. 332,264.

Kramer,C.et. al.,1998, Astron. Astrophys.329, 249.

Larson, R.B.,1981,MNRAS 194,809.

Larson, R.B., 1995 ,MNRAS 272, 213.

Malin, D., 1983, A View of the Universe, CUP.,pg.133.

Mandelbrot, B.B., 1982, The Fractal Geometry of Nature, Freeman, N.Y.

Larson, R.B., 1995 ,MNRAS 272, 213.

Paley, E., et.al.,1991, Ap.J. 376, 335.

Peitgen,  H.O., Jurgens,  H. and  Saupe,  D., 1992,  Fractals for  the
Classroom- Introduction to Fractals and Chaos, Springer-Verlag, N.Y.

Prasad,  S.S.   ,1995,  Fragmentation  of Molecular  Clouds  and  Star
Formation, eds.E.   Falgarone, F.   Boulanger and G.   Douvert, Kluwer
Academic, Dordecht, p. 93.

Silk, J., 1995, Ap. J. 438, L41.

Stark, R., 1995, Astron. Astrophys. 301, 873.

Stutzki, J., et. al., 1998, Astron. Astrophys. 336, 697.

van der Marel,R.P., et.al.,1997, Nature 385, 610.

Weber, F., 2000, astro-ph 0008376.

Williams, J.P., et. al., 1999, astro-ph 9902246.

White, G.J., et. al., 1997, Astron. Astrophys. 323, 931.

}

\vskip 20 pt
\centerline{\bf Acknowledgements}
\bigskip

\noindent

The author  wishes to  express her gratitude  to her  supervisor Prof.
B.Basu, former  Head of Department of  Applied Mathematics, University
of Calcutta,  for support and guidance  recieved througout; Dr.M.Hart,
Dept. of Pure Mathematics;  Dr.C. Tadhunter, Department of Physics and
Astronomy, University of Sheffield, UK;Prof.A. Boksenberg,Institute of
Astronomy, Cambridge,  UK; Prof. R.Gupta, IUCAA, for facilitating this
work; G.Selvakumar, M.Appakutty, Vainu  Bappu Observatory, Kavalur, for
expert  assistance  recieved   during  observations;  members  of  the
Telescope  Committee   at  the  Indian   Institute  for  Astrophysics,
Bangalore,  for  the observing  time  granted.   Thanks  also goes  to
Y.Wadekar,  IUCAA  and  C.D.   Ravikumar,  Dept.  of  Physics,  Cochin
University of  Science and Technology, India,  for assistance recieved
for the image reductions; Prof. Narlikar, D.Gadre and D.  Mitra,IUCAA,
for useful discussions.

\end